\documentclass[twocolumn]{revtex4-2}

\usepackage{amsmath}
\usepackage{graphicx}

\def\nn{\nonumber}

\begin{document}

\title{Stationary Two-State System in Optics using Layered Materials}

\author{Ken-ichi Sasaki}
\email{ke.sasaki@ntt.com}
\affiliation{NTT Research Center for Theoretical Quantum Physics and NTT Basic Research Laboratories, NTT Corporation,
3-1 Morinosato Wakamiya, Atsugi, Kanagawa 243-0198, Japan}

\date{\today}

\begin{abstract}
 In scenarios where electrons are confined to a flat surface, such as graphene, 
 quantizing electrodynamics reveals intriguing insights.
 We find that one of Maxwell’s equations manifests as part of the Hamiltonian, 
 leading to novel constraints on physical states due to residual gauge invariance. 
 We identify two quantum states with zero energy expectation values: 
 one replicates the scattering and absorption of light, a phenomenon familiar in classical optics,
 while the other is more fundamentally associated with photon creation.
 These states form an inseparable two-state system, 
 giving a new formula for reflection and transmission coefficients with photon emission effects. 
 Notably, there exists a special thickness of the surface where these states decouple, 
 offering intriguing possibilities for exploring physics through symmetry-based perturbations 
 involving concepts of parity, axial gauge fields, and surface deformation.
\end{abstract}

\maketitle

\section{Introduction}

Two-state systems represent the simplest cases that exhibit the essential features of quantum physics.
Examples include spin-$1/2$, the two base states for the ammonia molecule, 
(clockwise and counterclockwise) persistent currents of superconducting qubit, 
and the hydrogen molecular ion.~\cite{Feynman1989,Sakurai2017}
In optical science, the two-state polarization of the radiation field plays a fundamental role in 
uncovering the laws of physics, such as the no-cloning theorem,~\cite{Wootters1982} 
serving as a source of quantum entanglement,~\cite{Kocher1967,Aspect1982,Ou1988,Takesue2004} 
and contributing to modern optical transmission technology, 
allowing for the simultaneous transmission of double the amount of information.
Recently, the two-state coherent phase of a pulsed field has been related to an artificial spin.~\cite{Wang2013,Honjo2021}
There may be more two-state systems in optics beyond these examples.
Even when a system appears to have many degrees of freedom, 
it is sometimes apparent that, from a certain perspective,
only two crucial states play an essential role in the mechanism of the phenomenon
under investigation.~\cite{Bohm1951,Pines1952,Sasaki2006c,Sasaki2008e}

In this paper, we employ quantum electrodynamics~\cite{sakurai67}
to illustrate the existence of two stationary quantum states in optics involving layered materials.
Our formulation reveals that the appearance of these two states is 
intricately connected to gauge invariance,
and the origin of the two-state degree of freedom is inherently linked to the 
chiral nature of photons, specifically the right- and left-going components.
A general formula for the reflection (or transmission) coefficient corrected by photon emission is presented.
Moreover, we show that the antisymmetric combinations of counter-propagating photons are 
connected to the axial gauge field, suggesting a potential link to 
the geometrical deformation of layered materials and the Raman effect.

This paper is structured as follows: 
In Sec.~\ref{sec:sec2}, 
we present fundamental insights into the quantum electrodynamics of layered materials.
By utilizing the residual gauge degrees of freedom, 
we obtain quantum mechanical conditions for physical states.
In Section~\ref{sec:sec3}, we construct two states that satisfy these conditions,
allowing us to derive a formula for the reflection (or transmission) coefficient corrected by photon emission.
Finally, Section~\ref{sec:sec4} discusses the findings, 
and Section~\ref{sec:sec5} concludes the paper.

\section{Local Hamiltonian and residual gauge symmetry}\label{sec:sec2}

The space is divided into two regions, $x<0$ (left) and $x>0$ (right),
by an absorbing surface located at $x=0$.
Some fraction of a continuum incident light coming from $x=-\infty$
is absorbed only by the surface where the electronic current ${\bf j}$ exists, 
and the remaining is either reflected back to $x=-\infty$ or transmitted forward to $x=+\infty$.
In the classical theory of optics, 
the reflection and transmission coefficients are derived from
Amp\`ere's circuital law, 
given by $c^2 \nabla \times {\bf B} = \frac{\bf j}{\epsilon_0} + \dot{\bf E}$.
Integrating this equation with respect to an infinitesimal element around the surface,
the magnetic field experiences a discontinuity at the surface,
expressed as 
$c^2 \{ B_y(0_+)-B_y(0_-) \}= \int_{0_-}^{0_+} \frac{j_z}{\epsilon_0} dx$,
due to the localization of the current 
$j_z = J_z \delta(x)$ about the surface.~\cite{Sasaki2020a,Sasaki2020b}
Meanwhile, the electric field $E_z$ remains continuous at the surface,
and the last term of Amp\`ere's circuital law vanishes upon integration.
Since the magnetic field is expressed in terms of the radiation gauge field 
by $B_y=-\partial_x A_z$, we obtain the boundary condition
\begin{align}
 \partial_x A_z|_{x=0_+} - \partial_x A_z|_{x=0_-} + \frac{J_z}{\epsilon_0 c^2}=0,
 \label{eq:bc}
\end{align}
showing that the first derivative of $A_z(x)$ is discontinuous at the surface,
meanwhile, $A_z(x)$ must be continuous.
Equation~(\ref{eq:bc}) is also derived as a characteristic of stationary states
within the Poynting's theorem.~\cite{Feynman1989}
The continuity equation is expressed as $E_z j_z + \partial_x S_x + \dot{U}_{em} = 0$,
where $S_x = -\epsilon_0 c^2 E_z B_y$ represents the Poynting vector, and 
$U_{em} = \frac{\epsilon_0}{2} (E_z^2 + c^2 B_y^2)$
is the energy density of the electromagnetic field.
In the context of stationary states, $\dot{U}_{em}=0$, allowing us to reproduce Eq.~(\ref{eq:bc})
by integrating $E_z j_z + \partial_x S_x = 0$ with respect to an infinitesimal element around $x=0$.
Now, in quantum field theory,
$A_z(x)$ as well as $J_z$ is a field operator consisting of particle creation and annihilation operators.

It is possible to show that Eq.~(\ref{eq:bc}) appears as a part of Hamiltonian
and is related to the gauge invariance in the fundamental level.
For the photon's Hamiltonian of
\begin{align}
 H_{ph} = \frac{\epsilon_0}{2} \int_{-\infty}^{+\infty} dx \left\{ \dot{A}_z(x)^2 + c^2 ( \partial_x A_z(x))^2 \right\},
\end{align}
the integral may be divided into a left region, right region,
and the surface $[0_-,0_+]$.
Then, the Hamiltonian of the radiation field at the surface is given by 
$\epsilon_0 c^2 A_z(x) \partial_x A_z(x) |_{0_-}^{0_+}$.
By combining it with the gauge coupling of the interaction Hamiltonian, $H_{int} = A_z(0) J_z$,
we obtain the Hamiltonian at the surface $H_{ph}|_{x=0}+H_{int} (\equiv H_{local})$ as
\begin{align}
 \epsilon_0 c^2 A_z(0) 
 \left\{ \partial_x A_z(x)|_{x=0_+} - \partial_x A_z(x)|_{x=0_-} + \frac{J_z}{\epsilon_0 c^2} \right\}.
\end{align}
To simplify the notation, we use ${\cal B}$ to denote the operator of the boundary condition, 
${\cal B} \equiv \partial_x A_z(x)|_{x=0_+} - \partial_x A_z(x)|_{x=0_-} + \frac{J_z}{\epsilon_0 c^2}$,
and the local Hamiltonian is rewritten as 
\begin{align}
 H_{local}=\epsilon_0 c^2 A_z(0){\cal B}.
\end{align}
The energy must be invariant with respect to a residual gauge transformation,
given by $A_z(0) \to A_z(0) + \partial_z \lambda(x,z)|_{x=0}$, where $\lambda(x,z) = C z$ satisfies
$\partial^2_z \lambda(x,z)=0$.
However, the local Hamiltonian changes under such a residual gauge transformation as
$H_{local} \to H_{local} + \epsilon_0 c^2 C{\cal B}$.
Thus, to ensure that a residual gauge transformation does not alter the energy of physical states, 
any physical state $|\Phi\rangle$ must satisfy the condition $\langle \Phi| {\cal B}| \Phi\rangle=0$.
Instead of stating that Eq.~(\ref{eq:bc}) is satisfied as the expectation value (Ehrenfest's theorem),
Eq.~(\ref{eq:bc}) is satisfied as a consequence of the residual gauge symmetry.~\cite{Kugo1979}
Additionally, any physical state should not overlap with a state 
generated by acting the vacuum with ${\cal B}$, i.e., 
$\langle \Phi| {\cal B}\left( | 0 \rangle \otimes |vac\rangle \right)=0$, where 
the ground state of the photon (matter) is expressed by $|0\rangle$ ($|vac\rangle$).
Failure to meet this condition could allow a residual gauge transformation to introduce an arbitrary amount of energy,
which is a scenario not observed in nature.
We note that 
the result of the Poynting's theorem, $E_z {\cal B} = 0$, can be derived from $\dot{H}_{local}=0$
and $\dot{\cal B} = 0$, because $E_z = - \dot{A}_z$.

\section{Construction of two quantum states}\label{sec:sec3}

\begin{figure}[htbp]
 \begin{center}
  \includegraphics[scale=0.5]{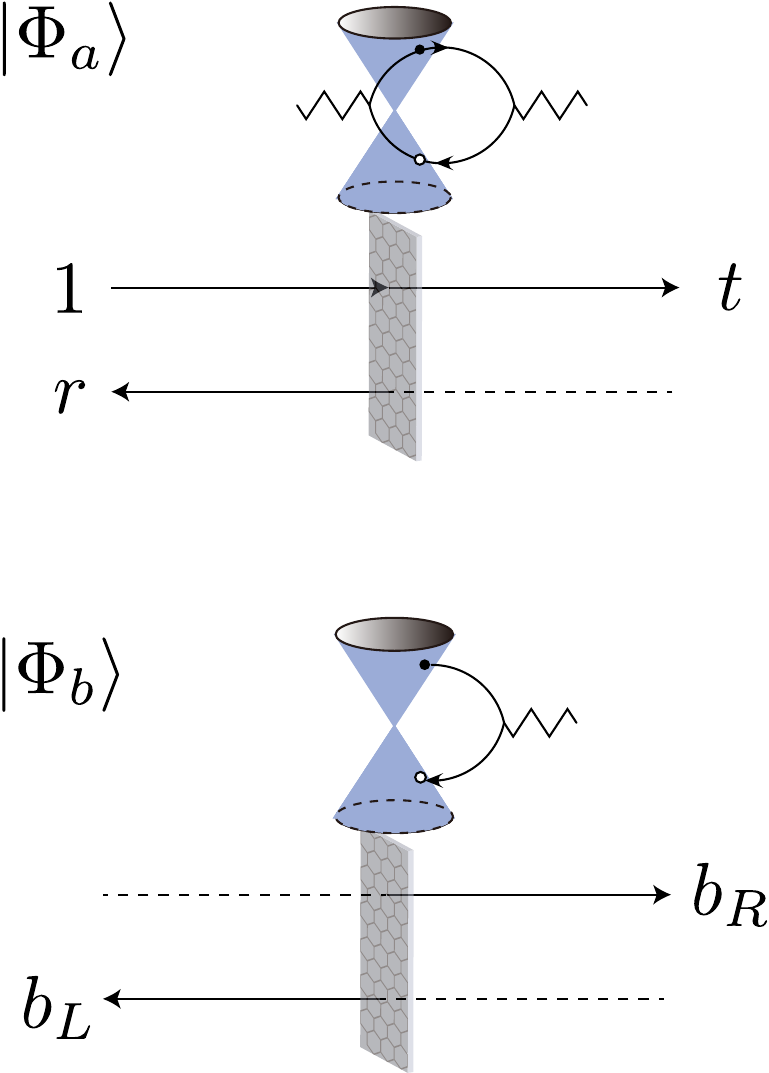}
 \end{center}
 \caption{Schematic of the two-state system:
 It consists of $|\Phi_a \rangle$ and $|\Phi_b \rangle$, 
 having a zero energy expectation value of $H_{local}$ due to the residual gauge symmetry.
 (Above) The arrow with ``1'' from left to right expresses the incident light ray, 
 which is reflected or transmitted by a thin film (graphene).
 The dashed line represents the zero photon state.
 (Below) The state of light emission from the excited state is symmetrical in left and right directions.
 Only the off-diagonal matrix element of $H_{local}$, $\langle \Phi_b | H_{local} | \Phi_a \rangle$, 
 is non-zero and proportional to $b_R^* t + b_L^* r$.
 The states of graphene are represented by the electron-hole pairs ($\bullet$ and $\circ$)
 in the Dirac cone (zigzag line denotes light).
 The electron-hole pairs in $|\Phi_a \rangle$ are virtual states of photons, 
 while those in $|\Phi_b \rangle$ are real states 
 that are constantly supplied by $|\Phi_a \rangle$ 
 (i.e., $\langle \Phi_b | \Phi_a \rangle \ne 0$).
 }
 \label{fig:1}
\end{figure}

We will show that the Hilbert space holds two quantum states
denoted by $|\Phi_a \rangle$ and $|\Phi_b \rangle$
(see Fig.~\ref{fig:1}).
Each state is not a single quantum but a specific combination of light and matter quanta.
The former reproduces scattering and absorption of light by matter 
which is consistent with the classical description of optics,
and the latter is more fundamentally related to the photon creation or light emission.
We use the Schr\"odinger picture (representation),~\cite{Sakurai2017,sakurai67} 
and the operators are time-independent.
Moreover, the state vectors we construct are stationary states with zero-energy of the local Hamiltonian,
the variable of time is concealed.
The gauge field operator is a sum of right- and left-moving components
as $A_z(x) = A_R(x) + A_L(x)$.
The electric field operator is denoted by $E_z(x)$
which forms canonical equal-time commutation relations with $A_z(x)$, 
such as $\left[ A_z(x),E_z(x') \right] = -i\frac{\hbar c}{\epsilon_0} \delta(x-x')$, 
$\left[ A_R(x),E_R(x')  \right] = -i\frac{\hbar c}{2\epsilon_0} \delta(x-x')$, and 
$\left[ A^{+}(x),E^{-}(x')  \right] = -i\frac{\hbar c}{2\epsilon_0} \delta(x-x')$, 
where $-$ and $+$ of the superscript stands for 
the annihilation and creation part of the field operators, respectively.
Note that $A_z(x)$ commutes with $B_y(x')$ and so with ${\cal B}$.
However, each component satisfies
$[\partial_x A_R(x),A_R(x')]= +i\frac{\hbar}{2\epsilon_0} \delta(x-x')$ and 
$[\partial_x A_L(x),A_L(x')]= -i\frac{\hbar}{2\epsilon_0} \delta(x-x')$,
which are canceled in
$[B_y(x),A_z(x')]= 0$.

First, we define an entangled state
that is a superposition of right- and left-going photons
counter-propagating along the $x$-axis.
The spatially uniformity of the light state is perturbed by the absorbing surface as
$|a \rangle = N_a \int_{-\infty}^{+\infty} a(x)dx |0\rangle$ and
\begin{align} 
 a(x) = \left[ \theta(-x) + t \theta(x) \right] e^{i\frac{\omega}{c}x} A_R(x)
 + r \theta(-x) e^{-i\frac{\omega}{c}x}A_L(x).
\end{align}
Here, $t$ ($r$) is the transmission (reflection) coefficient, 
$N_a$ is a normalization constant of the state $|a\rangle$ (space is taken to be infinity),
and $\theta(x)$ is the step function; $\theta(x)=1$ for $x>0$ and vanishes otherwise.
We show below that 
the following quantum state reproduces the result obtained in classical optics,~\cite{Shen2005}
\begin{align}
 |\Phi_a \rangle \equiv |a \rangle \otimes |vac\rangle 
 + \frac{\langle 0| E_z(0) |a\rangle}{g} |0\rangle \otimes J_z|vac \rangle.
\end{align}
Since the electric field is continuous at the surface (as we see below), 
we can also take either $x=0_+$ or $0_-$ for the argument of $E_z$.
A constant $g$ is introduced for dimensional consistency whose value is not of particular importance here.
The classical physics is reproduced by replacing $|a\rangle$ with a coherent state
$|A \rangle = N_A e^{i\int_{-\infty}^{+\infty} a(x) dx}|0\rangle$.

To specify $|\Phi_a \rangle$ uniquely, 
we use two physical conditions.
First condition is the continuity of the electric field 
$\langle 0 | E_z(0_-) |a \rangle = \langle 0 | E_z(0_+) |a \rangle$
giving $1 + r = t$.
Second is the necessary condition with respect to a physical state,
\begin{align}
 \langle 0|\otimes \langle vac| {\cal B} |\Phi_a \rangle = 0.
 \label{eq:physc}
\end{align}
Using commutation relations, we obtain from Eq.~(\ref{eq:physc}) [see Appendix~\ref{app:1}]
\begin{align}
 i \frac{\hbar}{2\epsilon_0} t -i \frac{\hbar}{2\epsilon_0} (1-r) 
 + \frac{\sigma g}{\epsilon_0 c^2}i \frac{1}{g} \frac{\hbar c}{2\epsilon_0} t= 0.
 \label{eq:classicalformula}
\end{align}
We note that 
$\langle vac|J_zJ_z|vac\rangle = g \sigma$, 
where $\sigma$ is dynamical conductivity (or vacuum polarization).
In the case of graphene, the dynamical conductivity is given by $\sigma=\pi \alpha \epsilon_0 c$,
where $\alpha$ is the fine-structure constant
(the absorption probability is $\pi \alpha \sim$ 2.3 percent).~\cite{Shon1998,Nair2008,ando02-dc}
Thus, these two conditions lead to
$t = \frac{2}{2+\pi \alpha}$ and $r = \frac{-\pi \alpha}{2+\pi \alpha}$.
Moreover, we obtain $\langle \Phi_a| H_{local}  |\Phi_a \rangle = 0$ which 
also reproduces the energy (or probability) conservation given by Poynting vector, 
$1 - r^2 - t^2 = \pi \alpha t^2 (\equiv A_{abs})$.~\cite{Sasaki2020a,Sasaki2020b}

Next we construct the other state vector,
\begin{align}
 |\Phi_b \rangle 
 \equiv |b \rangle \otimes |vac\rangle 
 + \frac{\langle 0| E_z(0) |a\rangle}{g}  |0\rangle \otimes |exc\rangle,
\end{align}
where $|b\rangle =N_a \int_{-\infty}^{+\infty} b(x) dx|0 \rangle$
and $b(x) = b_{R} \theta(x) e^{i\frac{\omega}{c}x} A_R(x) + b_{L} \theta(-x) e^{-i\frac{\omega}{c}x} A_L(x)$.
It represents light emission from the surface.
The presence of $\langle 0| E_z(0) |a\rangle$ on the last term 
shows that the emission is the spontaneous one (stimulated one when $|a\rangle \to |A\rangle$).
The state vector of the matter $|exc\rangle$ denotes the excited state.
This state is specified primarily by $\langle vac | J_z|exc \rangle =g F$.
In the same way as $|\Phi_a \rangle$,
the continuity of the electric field with respect to $|b \rangle$, 
$\langle 0 | E_z(0_+) |b \rangle = \langle 0 | E_z(0_-) |b \rangle$,
and the physical state condition $\langle 0|\otimes \langle vac| {\cal B} |\Phi_b \rangle = 0$
are sufficient to obtain that $b_R = b_L (\equiv b)$ and 
\begin{align}
 b =  -\frac{F}{2\epsilon_0 c} t.
\end{align}
This state automatically satisfies $\langle \Phi_b| H_{local} |\Phi_b \rangle = 0$.
Thus, we found that absorption means the two state-vectors
and that the strength of emitted light depends on the electric field at the graphene layer ($t$).

Since $J_z^{+}J_z^{-}$ contains the number operator of electron-hole pairs $N_{eh}$, 
the expression $g^2 |F|^2= g\sigma \langle exc|N_{eh}|exc \rangle$ is derived, 
representing the proportionality to the absorbed photon energy.
Thus, the magnitude of $F$ is expressed as $|F| \propto \sqrt{A_{abs}}$, 
with the sign of $F$ depending on the specific characteristics of the matter system.~\cite{Durrant1998,Cray1998}
Let's introduce a phenomenological parameter ${\cal B}$
which gives the emission-to-absorption ratio.
It leads to the expression
$|F({\cal B})| = \epsilon_0 c \sqrt{2{\cal B} A_{abs}}$ or $|b({\cal B})| = \sqrt{({\cal B}/2) A_{abs}} |t|$.
It is desirable to determine the value of ${\cal B}$ theoretically, but is beyond the scope of this paper.
However, it's worth noting that ${\cal B}$ may be influenced by the surrounding environment;
for example, 
when graphene is suspended in the air, 
lattice vibrations do not decay into substrates,
potentially reducing the relative significance of the paths from the excited states to light emission.

\begin{figure}[htbp]
 \begin{center}
  \includegraphics[scale=0.5]{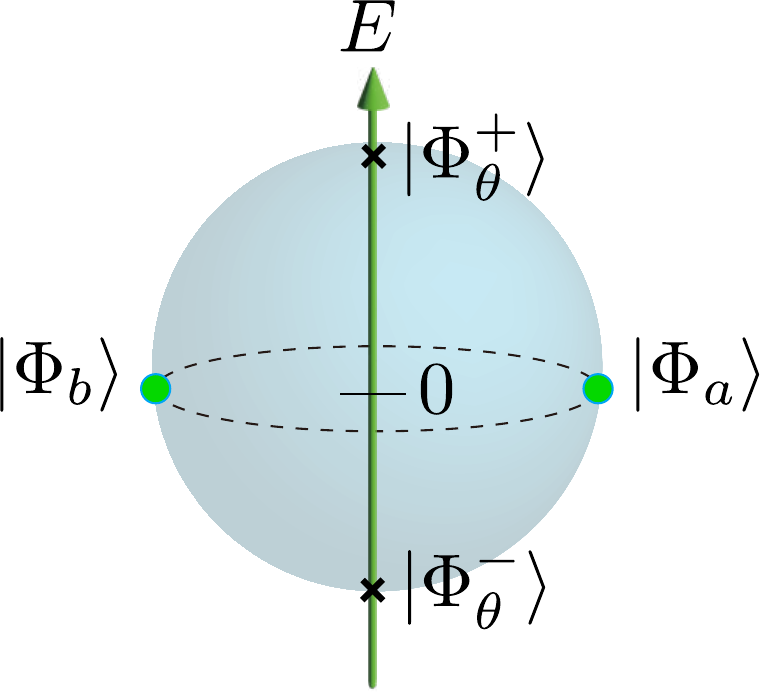}
 \end{center}
 \caption{Schematic of the two-state system in a ``sphere'':
 The vertical axis represents the energy expectation value of $H_{local}$. 
 The equator corresponds to zero-energy states, including $|\Phi_a \rangle$ and $|\Phi_b \rangle$.
 The north (south) pole corresponds to $|\Phi_\theta^+ \rangle$ ($|\Phi_\theta^- \rangle$).
 Each point of the surface of the sphere corresponds to a quantum state.
 In Ref.~\onlinecite{Sasaki2024a}, we observed the south pole
 and accessed to $|\Phi_b \rangle$.
 As $t+r\to 0$ (or $N\to 2/\pi\alpha$), the north and south poles approach each other along the vertical axis.
 }
 \label{fig:2}
\end{figure}

Though the zero-energy conditions $\langle \Phi_a |H_{local}|\Phi_a \rangle=0$
and $\langle \Phi_b |H_{local}|\Phi_b \rangle=0$ are both satisfied,
the off-diagonal element $\langle \Phi_b |H_{local}|\Phi_a \rangle$ is generally non-zero as
\begin{align}
 \langle \Phi_b | H_{local} |\Phi_a \rangle 
 = \epsilon_0 |E_{in}|^2 \tilde{\varepsilon} b^*(t+r),
\end{align}
and therefore the two states evolve inseparably by the Hamiltonian.
Here, $E_{in} \equiv \langle 0 |E_R(-\infty)|\Phi_a \rangle$
and $\tilde{\varepsilon}={\cal O}(1)$
[see Appendix~\ref{app:1}].
We shall introduce two states written as 
$|\Phi_\theta^\pm \rangle = |\Phi_a \rangle \pm e^{i\theta}|\Phi_b \rangle$,
where the relative phase $\theta$ is determined so that these states are decoupled;  
$ \langle \Phi^+_\theta | H_{local}  |\Phi^-_\theta \rangle = 0$ 
or ${\rm Im}[e^{i\theta}(t+r)^* b]=0$.
Thus, $e^{i\theta} = \pm \frac{t+r}{|t+r|} \frac{b^*}{|b|}$.
Then, their energy expectation values become non-zero as 
$\langle \Phi_\theta^\pm |H_{local}| \Phi_\theta^\pm \rangle = \pm 2(\epsilon_0 |E_{in}|^2 \tilde{\varepsilon}){\rm Re}[b^*(t+r)]$.
The relationship between these states is illustrated in Fig.~\ref{fig:2}.

An analogy to the electron spin is useful in understanding the physics.
Suppose we have up and down spin states, $|\uparrow\ \rangle $ and $|\downarrow\ \rangle$,
which are the eigenstates of $\sigma_z$.
Let us assume that each spin state is in a magnetic field along the $y$-axis.
The energy expectation value vanishes.
Now, $|\Phi_\theta^\pm \rangle$ corresponds to 
$|\pm \rangle \equiv (|\uparrow\ \rangle \pm e^{i\theta} |\downarrow\ \rangle)/\sqrt{2}$.
When $\theta$ is determined so that $\langle + |\sigma_y | - \rangle = 0$,
we have $\langle \pm |\sigma_y | \pm \rangle = \pm 1$.
Note that the general expression of $H_{local}$ in terms of the spin is 
$H_{spin}=a_+ \sigma_+ + a_- \sigma_-$, where $\sigma_\pm = \sigma_x \pm i \sigma_y$ is the ladder operators
and $a_\pm$ is the amplitudes.
It satisfies $\langle \ \uparrow |H_{spin}|\uparrow\ \rangle =0$ and $\langle \ \downarrow|H_{spin}|\downarrow\ \rangle =0$.

The reflection coefficient is calculated as
$\langle 0 |E_L(-\infty)|\Phi^\pm_\theta \rangle$,
where $|\Phi_b \rangle$ emerges as a correction to the reflection coefficient.
There exist two possibilities for the corrected reflection coefficient, represented as:
\begin{align}
 r_\pm({\cal B}) \equiv r \pm e^{i\theta} b = r \pm \left( \frac{t+r}{|t+r|} \right) |b({\cal B})|.
\end{align}
When the lower energy state $|\Phi^-_\theta \rangle$ is realized,
the actual reflection is given by $r_-({\cal B})$.
This formula for the reflection coefficient holds significant experimental relevance.
Indeed, we recently applied this reflection formula 
to analyze experiments conducted on graphene placed on substrates.~\cite{Sasaki2024a}
Our findings show that for visible range, $r^-({\cal B})$ with ${\cal B} = 0.0007$ 
reproduces the experimental results quite accurately.
Additionally, we discovered that one can access $|\Phi_b \rangle$
by suppressing the contribution of $|\Phi_a \rangle$ to the reflection ($r \sim 0$)
using multilayer graphene (about 20 layers) and exploiting the 
destructive interface effect of 
SiO$_2$/Si substrates.~\cite{Sasaki2024a}

Due to absorption, the two states slightly overlap, i.e.,
$\langle \Phi_b | \Phi_a \rangle \ne 0$, and 
a simple interpretation of a two-state system is difficult to be applied.
Apparently, from the reflected (or transmitted) light we will observe, 
the signal is inseparable between $|\Phi_a \rangle$ and $|\Phi_b \rangle$.
If the two states were truly indistinguishable from each other, 
the situation is something like an electron spin in the absence of a magnetic field, and 
we need to face the issue of control.
However, at least when $t+r=0$ is satisfied, $\langle \Phi_b |H_{local}|\Phi_a \rangle$ vanishes.
Then, we define new base states,
$|\Phi_A \rangle = |\Phi_a \rangle$
and $|\Phi_B \rangle = |\Phi_b \rangle-|\Phi_a \rangle \langle \Phi_a | \Phi_b \rangle$,
which become orthogonal and are not mixed by the local Hamiltonian.
Such an interesting situation is indeed possible to realize 
for multilayer graphene.~\cite{Sasaki2020a}
Let the layer number is $N$.
The corresponding transmission and reflection coefficients 
are given by the replacement $\pi \alpha \to N \pi \alpha$ 
in $t = \frac{2}{2+\pi \alpha}$ and $r = \frac{-\pi \alpha}{2+\pi \alpha}$.
Thus, $t+r=0$ when $N=2/\pi \alpha$.

\section{Discussion}\label{sec:sec4}

Perturbations should be incorporated into $H_{local}$,
as they have the potential to disrupt stabilization and counteract corrections through interference.
These perturbations arise from lattice vibrations and electron-phonon interactions,
particularly in relation to the Raman effect.
In this context, we examine some of these perturbations on symmetry considerations.

Firstly, we observe that $H_{local}$ remains invariant under parity,
as the fields transform according to
$A_z(x)\to -A_z(-x), B_y(x) \to B_y(-x)$, and $J_z \to -J_z$.
Let us introduce the axial gauge field $A^{ax}_z \equiv A_R-A_L$,
which is even under parity as $A^{ax}_z(x)\to A^{ax}_z(-x)$.~\cite{bertlmann00}
This is in contrast to the polar one $A_z (=A_R+A_L)$.
Since $A_z(x)$ and $A_z^{ax}(x)$ are regarded as the bonding and anti-bonding orbitals of 
counter-propagating photons, $A_z^{ax}(x)$ must be useful in discussing energy stabilization
governed by graphene.
Secondary, we point out that
the current $J_z$ in graphene comprises two currents of different valleys $K$ and $K'$,
as $J_z = J^K_z + J^{K'}_z$, which is even under the interchange of $K$ and $K'$.
It is known that the axial current $J^{ax}_z$, which is odd under the valley degrees of freedom
as $J^{ax}_z = J^K_z - J^{K'}_z$, couples to a field $A_z^{q}(0)$
expressing distortions in graphene's hexagonal lattice.~\cite{Sasaki2006c,Sasaki2008e}
For example, the primary Raman band in graphene, the $G$ band, results from their (phonon-electron) coupling:
$A_z^{q}(0) J^{ax}_z$.

When the system is invariant under parity transformation,
the perturbation in the form of $A_z(0)J^{ax}_z$ is allowed, 
but $A_z^{ax}(0)J^{ax}_z$ and $A^{ax}_z(0)J_z$ are not.
These are possible perturbations that make
the system not invariant under parity transformation.
Thus, there is a potential connection between the Raman effect and the axial gauge field,
as $J_z^{ax}$ can act as the source of $A_z^{ax}$ and $A_z$.
To further elucidate this point, 
we rewrite the photon wavefunctions in terms of $A_z$ and $A_z^{ax}$ as
\begin{align}
 a(x) &=  \frac{e^{i\frac{\omega}{c}x}}{2}(A_z(x) + A^{ax}_z(x)) \nn \\
 &+ \frac{r}{2} \left[ e^{i\frac{\omega}{c}x} \theta(x) + e^{-i\frac{\omega}{c}x} \theta(-x) \right]  A_z(x) \nn \\
 &+ \frac{r}{2} \left[ e^{i\frac{\omega}{c}x} \theta(x) - e^{-i\frac{\omega}{c}x} \theta(-x) \right]  A^{ax}_z(x), \\
 b(x) &= \frac{1}{2}\left[ b_R e^{i\frac{\omega}{c}x} \theta(x) + b_L e^{-i\frac{\omega}{c}x} \theta(-x) \right] A_z(x) \nn \\
 &+ \frac{1}{2}\left[ b_R e^{i\frac{\omega}{c}x} \theta(x) - b_L e^{-i\frac{\omega}{c}x} \theta(-x) \right] A^{ax}_z(x).
\end{align}
Since wavefunctions are continuous, $A^{ax}_z(x)$ must vanish at the surface ($x=0$) for both $a(x)$ and $b(x)$.
However, $b(x)$ allows a nonzero $A^{ax}_z(x)$ if $b_R = -b_L$.
Thus, if graphene can excite $A^{ax}_z(x)$, the effect appears as a special light emission,
whose wavefunction is orthogonal to $|\Phi_b \rangle$.
The light emission is a quantum effect because $A^{ax}_z(x)$ satisfies
$[B_y(x),A^{ax}_z(x')]= -i\frac{\hbar}{\epsilon_0} \delta(x-x')$
and $[B^{ax}_y(x),A_z(x')]= +i\frac{\hbar}{\epsilon_0} \delta(x-x')$, where $B_y^{ax}(x) \equiv \partial_x A_z^{ax}(x)$
(the latter is what we have used for the anomalous commutator).
In the spin analogy, these perturbations are akin to those proportional to $\sigma_x$ (that cause a change in $\theta$)
and $\sigma_z$ (that mixes $| + \rangle$ and $|-\rangle$).

The above discussion on phonons might be important when combined with the residual gauge symmetry of light. 
For example, when examining a dynamical transition (between two states $|\Phi_\theta^\pm \rangle$) 
induced by certain time-dependent perturbations, such as forced lattice vibrations, 
the memory effects arising from the system’s history due to coupling with the environment become significant.
Such memory effects are not accounted for in standard Markov approximations.~\cite{Breuer2007,Breuer2016}
Recently, Shen {\it et al.} examined non-Markovian effects within the framework of waveguide quantum electrodynamics,~\cite{Shen2023,Li2024}
aiming to explore the generation of multiple complex single-photon wavepackets from an optical cavity 
coupled to driven three-level atoms with non-Markovian input-output fields.

The application of the theoretical framework developed for a surface
to $N$-layer is straightforward,
when the electronic current flowing between layers is negligible.
The total Hamiltonian is a sum of the local Hamiltonian as
$H_{total} = \sum_{j=1}^N  \epsilon_0 c^2 A_z(x_j){\cal B}(x_j)$.
The solution is built as a superposition of $|\Phi_{a_N} \rangle$ and $|\Phi_{b_j} \rangle$, 
which can be known using a transfer matrix method.~\cite{Sasaki2024a}
The reflectance is given by $R_N = |r_N + \sum_{j=1}^N e^{i\theta_j} b_j|^2$, 
where $\theta_j$ includes the phase of $\pm$ for the two energy levels.
In terms of the spin analogy, the problem is approached as a spin-chain.
This viewpoint may be useful in analysis of real systems since unexpected perturbations can exist.
For example, when perturbations flip the spin of a layer,
the reflection and transmission coefficients (near the layer) undergo changes, 
and $|\Phi_{a_N} \rangle$ is consequently modified.
This modification results in correlations among spins at different layers.
This aspect appears to be another important issue connected to super-radiance~\cite{Dicke1954,Sheremet2023}
and the synchronization of weakly coupled oscillators.~\cite{Kuramoto1984}

It is also straightforward to include the degrees of freedom of photon polarization.
Namely, we will have additional two states for $A_y$ as well as $A_z$.
Thus, the system holds four states in total.
Because the characteristics of the matter system enters through the current operator only,
our formulation is applicable to various layered systems besides graphene 
by taking the corresponding current operator $J$.
A fundamental issue is that $A_y$ is just a copy of $A_z$, or 
there is some sort of correlation between them.
In the case of pristine graphene, the current operators $J_y$ and $J_z$
excite different electron-hole pairs and therefore
such a correlation is suppressed by optical selection rule.~\cite{Sasaki2011a}

\section{Conclusion}\label{sec:sec5}

In conclusion, we have formulated a two-state system 
that arises from the distinct states of matter in layered materials: $J_z|vac\rangle$ and $|exc\rangle$.
Considering that photon states directly emerge from constraints imposed by residual gauge invariance, 
we propose that these two states of matter can be effectively 
characterized by distinct configurations of the light field.
The state $|\Phi_a \rangle$ appears asymmetrical with respect to $A_R$ and $A_L$,
yet it is solely described by the polar gauge field $A_z$.
On the other hand, $|\Phi_b \rangle$ while symmetrical with $b_R = b_L$, 
may incorporate the axial gauge field $A_z^{ax}$, which holds significance
in phenomena such as the Raman effect.
In the absence of perturbations,
the two-state system is degenerate, particularly 
for multilayer graphene with special layer numbers
$N \simeq 2/\pi \alpha$, where 
the matrix element of $\langle \Phi_b|H_{local}|\Phi_a \rangle$ vanishes.
Then, $|\Phi_a \rangle$ exhibits a perfect asymmetrical combination of $t=1/2$ and $r=-1/2$,
and $|\Phi_b \rangle$ exhibits a perfect symmetrical combination of $b_R$ and $b_L$.
Therefore, it is reasonable to understand that 
detecting a left-going or right-going mode does not provide information
about which state, $|\Phi_a \rangle$ or $|\Phi_b \rangle$, is realized.

\section*{Acknowledgments}

The author thank T. Matsui (Anritsu Corporation), M. Kamada (Anritsu Corporation),
and K. Hitachi for helpful discussions.

\appendix

\section{Derivation of Eq.~(\ref{eq:classicalformula})}\label{app:1}

The first term of Eq.~(\ref{eq:classicalformula}) is obtained by the following manipulations:
\begin{align}
 &\langle 0| \partial_x A_z(x)|_{x=0_+} |a \rangle 
 = N_a \langle 0| \partial_x A_z(x)|_{x=0_+} \int_{-\infty}^{+\infty} a(y) |0\rangle dy \nn \\
 &= N_a t \int_{0}^{+\infty} \theta(y) 
 e^{+i\frac{\omega}{c}y} \langle 0| \partial_x A_z(x)|_{x=0_+} A_R(y)|0\rangle dy \nn \\
 &= N_a t \int_{0}^{+\infty} 
 e^{+i\frac{\omega}{c}y} \langle 0| [\partial_x A_R(x)|_{x=0_+}, A_R(y)]|0\rangle dy \nn \\
 &= N_a \frac{i\hbar}{2\epsilon_0}t \int_{0}^{+\infty} 
 e^{+i\frac{\omega}{c}y} \delta(0_+-y) dy
 = N_a \frac{i\hbar}{2\epsilon_0}t, 
\end{align}
where $[\partial_x A_R(x),A_R(x')]= +i\frac{\hbar}{2\epsilon_0} \delta(x-x')$ is used.
The other terms of Eq.~(\ref{eq:classicalformula}) are obtained in a similar manner.
To demonstrate $\langle \Phi_a| H_{local}  |\Phi_a \rangle = 0$, we must compute terms 
such as $\langle a |A_z(0_+)\partial_x A_z(x)|_{x=0_+} |a \rangle$.
Since the operators $A_z(0_+)$ and $\partial_x A_z(x)|_{x=0_+}$ commute,
we treat it as $\langle a |A_z(0_+)|0\rangle \langle 0|\partial_x A_z(x)|_{x=0_+} |a \rangle$.
To calculate $\langle a |A_z(0_+)|0\rangle$,
we must consider the two-point correlation function 
such as $\langle 0 | [A_z(0_+ +\varepsilon),A_z(0_+)] |0 \rangle$, 
which exhibits an anomalous commutator part when the limit $\varepsilon\to 0$ is taken
(note that $0_+$ means the length scale of wave function spreading: thickness of graphene surface).
It is defined as 
$\lim_{\varepsilon \to 0}\langle 0 | [(A_R(x+\varepsilon)+A_L(x-\varepsilon)),A_z(x)] |0 \rangle$.
This yields:
$\lim_{\varepsilon \to 0} \varepsilon \langle 0 | [\partial_x A_R(x)-\partial_x A_L(x),A_z(x)] |0 \rangle 
= \lim_{\varepsilon \to 0}  \frac{i\hbar}{\epsilon_0}\varepsilon\delta(0)$, which is nonzero because 
$\lim_{\varepsilon \to 0}\varepsilon\delta(0)\equiv \tilde{\varepsilon}$ is finite.


%

\end{document}